# The Complex Irradiation Facility at DLR-Bremen

Thomas Renger[1], Maciej Sznajder[1, 2], Andreas Witzke[1] and Ulrich Geppert[1, 3]

*1. DLR Institute for Space Systems, System Conditioning, Bremen 28359, Germany*

*2. DLR Institute for Space Systems, University of Bremen, Bremen 28359, Germany*

*3. University of Zielona Góra, Kepler Institute of Astronomy, Zielona Góra 65-265, Poland*

**Abstract:** All material exposed to interplanetary space conditions are subject to degradation processes. For obvious reasons there is a great interest to study these processes for materials that are used in satellite construction. However, also the influence of particle and electromagnetic radiation on the weathering of extraterrestrial rocks and on organic and biological tissues is the research topic of various scientific disciplines. To strengthen the comprehensive and systematic investigation of degradation processes a new laboratory, the complex irradiation facility (CIF), has been designed, set up, tested, and put into operation at the DLR-Institute of Space Systems in Bremen (Germany). The CIF allows the simultaneous irradiation with three light sources and with a dual beam irradiation system for the bombardment of materials with electrons and protons having energies up to 100 keV. It is eminently suitable to perform a large variety of irradiation procedures that are similar to those which appear at different distances to the Sun. This paper is devoted to potential users in order to inform them about the capabilities of the CIF.

## 1. Introduction

The complex irradiation facility (CIF) was designed and commissioned with the aim to perform material investigations under radiation conditions as prevalent in space environment. The idea is to combine multiple radiation sources at a vacuum irradiation chamber. A driving motivation to establish the CIF at the DLR Institute of Space Systems was its participation in the DLR-ESA Gossamer solar sail project. However, the CIF can be used for a large variety of material studies. Besides of thin metallic foils, also thin sections of meteorites or organic substances can be exposed to a well-defined irradiation with protons, electrons, and electromagnetic waves. A great effort has been expended to simulate the conditions prevalent in the interplanetary space as realistic as possible. It concerns especially the dimensioning of the electron and proton accelerator with respect to energy and intensity range.

Both cover the bulk of the corresponding solar wind parameters. It concerns as well the quality of the vacuum which can be achieved in the CIF. To attain a high vacuum special effort was made. The quality of the vacuum is an important aspect to detect material degradation because molecules of the rest gas can interact with the radiation and degrade the surface of the sample. This would distort the results of degradation studies which are actually designed for conditions of the interplanetary space. Therefore the complete facility has been manufactured in UHV-technology with metal sealings and without organic compounds (rubber vacuum sealings, pump oil) to avoid self-contamination which could affect the results of the experiment.

It is known from many evaluation tests [1, 2] that particle and UV radiation can significantly degrade materials and, e.g., lead to changes in their mechanical behavior or thermo-optical properties [3]. That was taken into account for the determination of the required parameters of the CIF.

---

**Corresponding author:** Thomas Renger, scientist, research fields: material degradation and irradiation experiments. E-mail: Thomas.Renger@dlr.de.

An important number that characterizes the performance of the CIF is the acceleration factor. It is defined as the ratio of the intensity of the degrading radiation applied to a material during a laboratory test to the intensity of the same degrading parameter in space environment [2]. The spectra of the light sources are shown in comparison to the extraterrestrial solar spectrum [4, 5] in Subsections 3.1-3.3. The spectral distribution of the accelerating factor of the complete electromagnetic radiation provided by all light sources is presented in the conclusion.

Another aspect, beside the irradiation performance, is the measurement engineering to detect the changes of material properties caused by irradiation. For this purpose the CIF is equipped with a mass spectrometer at the irradiation chamber to qualify the outgassing behavior of the sample. Furthermore, our department has commissioned an optical measurement system, consisting of a FTIR spectrometer and Ulbricht spheres. It captures the changes of thermo-optical properties by measuring reflectivity and emissivity. This can be done presently only ex-situ, i.e., the sample has to leave the vacuum and the influence of the atmosphere cannot be avoided. However, it is foreseen until 2014 to achieve the development and commissioning of an in-situ measurement system. Another possibility to qualify optical properties is the measurement of the light pressure [6] before and after an irradiation test. Light pressure measurements return a kind of integral information about the state of degradation of irradiated metallic foils. Such a light pressure measurement facility will be placed outside the CIF.

In order to study the effects of degradation on to the elastic properties of the degraded materials, a tensile testing facility will be established at the CIF. Clearly, such studies can be performed only ex-situ.

The experimental studies of degradation effects have to be accompanied by theoretical efforts. This is necessary not only to get an idea about the physical processes that cause degradation. It is also necessary for profound predictions of changes in the material properties on long-term space missions, where a simple scaling by use of the acceleration factors provided by the CIF will fail. First attempts are in progress at DLR-Bremen [7].

## 2. Overview of the Configuration and Instrumentation of the CIF

### 2.1 Geometry and Technical Parameters

The CIF consists of a vacuum irradiation chamber (400 mm in diameter) which is connected to a lock chamber for the placement of the sample into the vacuum environment of the facility (Fig. 1). The irradiation chamber has got four tubes with flanges for the connection with the radiation sources. They are arranged at an angle of 30° to the neighboring one in the same level for the accelerator beam line, the solar simulator, and the argon-VUV-source. The Deuterium-UV-source is located above the solar simulator at an angle of 30° between their axes. The axes of all radiation sources are crossing in the center of the irradiation chamber where the target station is mounted. The geometry of that arrangement, the target mounting, and the radiation sources itself are dimensioned in a way, that a square area of 80 mm can be irradiated simultaneously with all sources.

The sample mounted in a holder will be inserted into the lock chamber and transferred after vacuuming by a magnetic manipulator into the sample station of the irradiation chamber. Additionally, there are magnetic manipulators installed in cross direction at the lock chamber. They can bring the sample holder in front of the Ulbricht spheres of the planned *in-situ* measurement system to measure the reflectivity before and after the irradiation without leaving the vacuum environment.

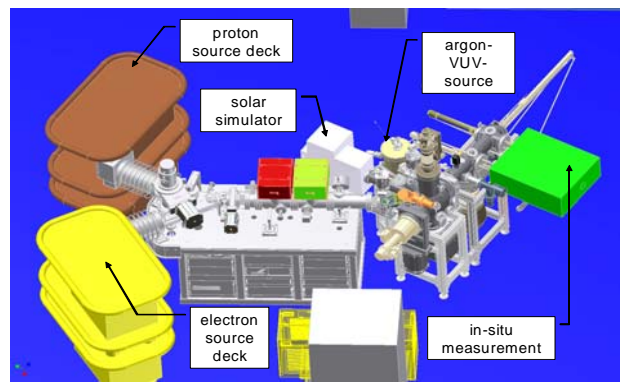

**Fig. 1   Schematic overall view of the CIF configuration.**

**Table 1   Technical parameters of the CIF.**

| Description | | Value |
|---|---|---|
| Vacuum irradiation chamber | Volume | Circa 33.5 l (400 mm diameter) |
| | Irradiated zone | 80 mm diameter |
| | Vacuum pressure | <10$^{-8}$ mbar (without VUV-source)<br><10$^{-6}$ mbar (depending on VUV settings) |
| Light sources | Solar simulator | 250 to 2,500 nm (5,000 W/m$^2$) |
| | Deuterium UV source | 112 to 400 nm (1.65 W/m$^2$) |
| | Argon VUV source | 40 to 410 nm (50 mW/m$^2$) |
| Proton and electron source | Current at lower energy range (1 to 10 keV) | 1 to 100 nA |
| | Current at higher energy range (10 to 100 keV) | 0.1 to 100 μA |
| Target thermal conditioning | Heating | Halogen radiation (600 W, 450 °C) |
| | Cooling | Liquid nitrogen (lN$_2$: 80 K) |

The vacuum system at the irradiation- and the lock chamber consists of a turbo molecular pump, an ion getter pump and a cryogenic pump to reach a pressure in the UHV-range. This will allow measurements with the quadrupole mass spectrometer which is installed at the irradiation chamber.

*2.2 Target Mounting*

The target mounting (Fig. 2) can be rotated about 30° into two directions. Thereby, either the VUV-radiation or the particle beam line can be perpendicularly oriented with respect to the sample surface. In the normal transfer position the sample surface is oriented perpendicular to the axis of the solar simulator.

The sample station is vertically adjustable, so that the sample holder can be inserted at the upper or the lower position at the station. The upper position provides a heating compartment from the backside of the sample. This can be used e.g., to simulate ambient conditions closer to the Sun, which cannot be supplied by the solar simulator. Furthermore the sample can be cooled to liquid nitrogen level to simulate ambient conditions of the deep space.

## 3. The Radiation Sources of the CIF

*3.1 The Solar Simulator*

The solar simulator is used to simulate the electromagnetic spectrum of the Sun in the wavelength range from 250 nm to 2,500 nm. It consists of a Xenon lamp with 1,600 W electrical power, an optical mirror and lens system to concentrate and homogenize the light spot. An air mass zero filter is mounted to eliminate the Xenon peaks and to fit the spectral distribution to the extraterrestrial one defined in the ASTM E-490 [4]. Fig. 3 illustrates the spectral irradiance of the solar simulator as measured at DLR-Berlin with the minimum, and the maximum settings of electrical power.

The measured values are corrected with the transmittance of the quartz glass window at the irradiation chamber. Presently there are ongoing activities to enhance the intensity in the UV range below 250 nm (Section 4).

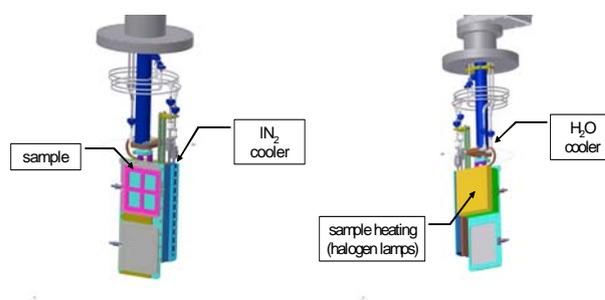

**Fig. 2   Sample station in the center of the irradiation chamber (front and back side).**

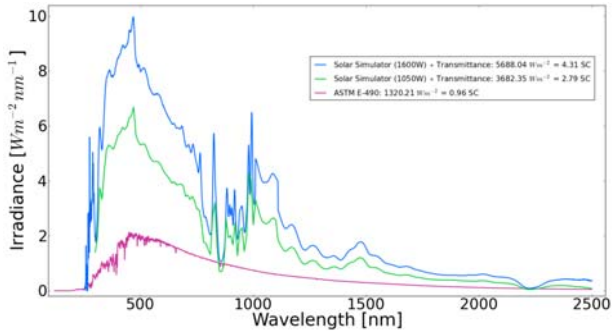

**Fig. 3  Spectral irradiance of the solar simulator with the minimum and maximum setting of electrical power in comparison to ASTM E-490 [4].**

### 3.2 The Deuterium-UV-Source

The deuterium-UV-source is a modified standard solution which is intended to be used as an excitation source in research and development. The modifications are the $MgF_2$-window that allows a transmission down to 112 nm, and an indium sealing between window and vacuum flange to avoid contamination.

The working principle of the source is the electrode excited gas discharge. The construction consists of a glass bulb, which is filled with deuterium, a cooling water jacket, and a vacuum flange (DN40CF). The electrical leads from the socket of the housing to the interior of the bulb are melt-sealed and vacuum tight. The glass bulb contains the thermionic cathode, which is heated by means of direct or alternating current, the ring-shaped anode, and an aperture of 1 mm diameter, where the discharge arc between cathode and anode is guided through. The light intensity can be controlled by the anode current between 0.8 A and 1.9 A. That corresponds to 120 W and 200 W electric power. The spectral irradiance, both at minimum and maximum setting, was calibrated by Physikalisch Technische Bundesanstalt (PTB) in Berlin (Fig. 4).

The stability of the source can be observed by shifting a monitoring sensor, mounted on a linear manipulator, onto the axis between source and sample.

### 3.3 The Argon-Gas-Jet-VUV-Simulator

The following description of the VUV-simulator is an abstract of the detailed presentation, its design, and calibration [8]. The advantage of the VUV-simulator is the generated electromagnetic radiation in a broad spectral range (soft X-rays, VUV, NUV) with relatively high intensities at lower wavelengths (< 110 nm). This is ensured by the design that avoids windows and is based on the semi-cryogenic version of a previous simulator, developed by the Institute of Low Temperature Physics and Engineering in Charkov, Ukraine [9].

The radiation is produced by the transition of electrons belonging to excited gas atoms into their ground state. The excitation occurs by electron bombardment of a spatially limited supersonic gas jet which flows into a vacuum chamber. Semi-cryogenic design means that the vacuum is maintained by a combination of cryogenic and mechanical pumps. Fig. 5 illustrates the arrangement of the electron source and the gas jet inside the VUV-chamber of the simulator.

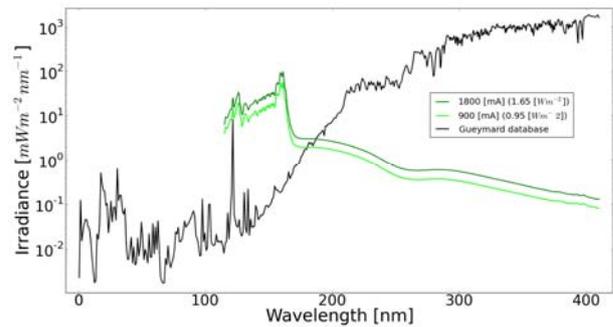

**Fig. 4  Spectral irradiance of the deuterium-UV-source with the minimum and maximum setting of electric power in comparison to Gueymard database [5].**

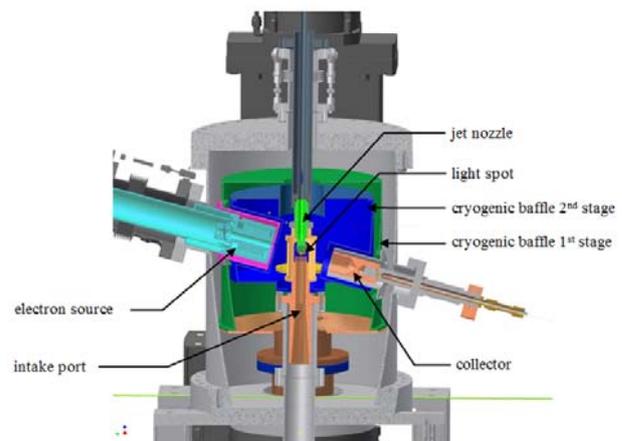

**Fig. 5  Sectioning of the VUV-simulator along the electron beam.**

The generated VUV-light is realized at the spot side perpendicular to the figure's plane. The electrons which pass through the gas jet are caught by the collector at the opposite site of the source. The components of the vacuum system are visualized better in Fig. 6, whose plane is perpendicular to the plane of Fig. 5.

The gas jet is injected by a nozzle from top of the VUV-chamber into the vacuum. The flow rate is stabilized by a flow controller and is adjustable by software in the range between 0 and 5,000 sccm (standard cubic centimeter per minute). The bulk of the gas load is pumped out through an intake port at the bottom of the VUV-chamber by a screw pump. The rest of the gas, which is a small fraction of about 3%, is removed by cryogenic condensation at two baffles. Each of them is connected to one of the both stages of the cold head of a commercial cryogenic pump (helium cooling machine). The 1st stage reaches a bottom temperature of 15 K without gas load. It increases to about 20 K under gas load by formation of Argon ice, which decreases the pumping power gradually and limits the operating time. The temperature is logged permanently as an indicator when regeneration is necessary to remove the ice. During this process the cryogenic pump is turned off while the mechanical pumps continue. The temperature inside the VUV-chamber increases and the residual gas, which was frozen on the cold baffles, is pumped out.

The gas nozzle is thermally connected to a water circuit for tempering it to avoid a freeze. A differential pumping segment is installed at the light outlet consisting of a turbomolecular pump and an aperture assembly. It improves the pressure conditions inside the irradiation chamber. So tests can be performed in the chamber at pressures of $10^{-6}$ mbar and less. The fraction of light which is permitted to enter the irradiation chamber is colored yellow in Fig. 6. An opening angle of about 6° is defined by the apertures in the cryogenic baffles and at the differential pumping segment. It ensures the irradiation of the target area with a diameter of 80 mm at a distance of about 770 mm from the spot. To avoid that charged particles produced at the spot can reach the irradiated object, an electrical filter is installed at the light outlet, which deflects them beyond the radiation flow.

The electron source is realized as a Pierce-type model with a magnetic lens behind the anode. The $LaB_6$-cathode is heated up electrically by adjusting the cathode voltage depending on the demanded emission current. The electronic control unit of the source provides a PID algorithm which stabilizes the emission current by varying the Wehnelt voltage automatically. The beam is focused by setting the current for the magnetically lens.

Two flanges are located at the opposite site of the light outlet. Each is connected with a window for visual inspection of the luminous jet and with a radiation indicator compartment to observe the stability of the radiation (Fig. 6).

The spectral intensity distribution of the radiation depends on the gas mixture, the flow rate of the jet, the electron current, on the alignment between beam and jet, and the focus of the electron beam. Fig. 7 illustrates these relations qualitatively by showing the size and intensity of the spot with different settings for the emission current and for the gas flow. The photos were taken through the window at the opposite site of the light outlet (Fig. 6).

During the calibration campaign at the radiometry laboratory of the PTB at Berlin Electron Storage Ring

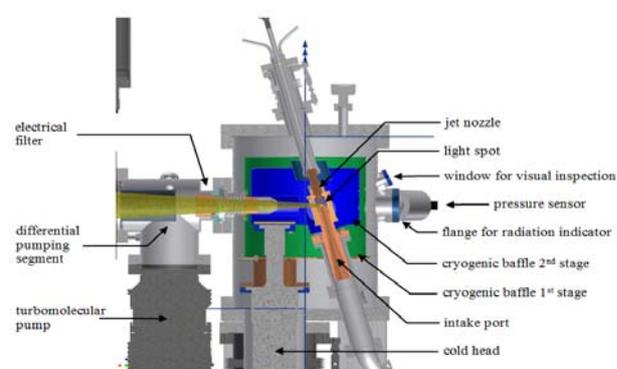

**Fig. 6 Sectioning of the VUV-simulator along the light outlet (yellow).**

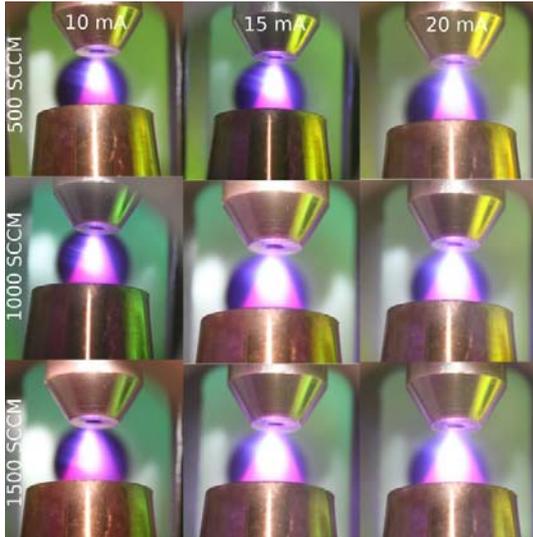

Fig. 7   Pictures of the VUV-spot with different settings of the gas flow rate (rows: 500, 1,000, 1,500 sccm) and the emission current (columns: 10, 15 and 20 mA).

Society (Bessy II) for Synchrotron Radiation in Berlin-Adlershof the alignment and the focus mentioned above were optimized. The following settings were chosen for the calibration:

The results of the calibration campaign are shown in Figs. 8 and 9 in the wavelength range from 40 nm to 410 nm.

*3.4 The 100 keV Proton/Electron Dual Beam Irradiation System*

The Dual Beam Irradiation System is designed to irradiate samples with protons or electrons independently or with both particle species simultaneously. The selected beam(s) can be scanned over the samples. All relevant parameters can be adjusted remotely via the computer control system. Fig. 10 illustrates the configuration of the beam line with its main items in view from top.

**Table 2   Parameters of the VUV-simulator for calibrated spectra.**

| Parameter | Value, Description |
| --- | --- |
| Gas mixture | Ar (98.5 %), He (0.5 %), Kr (1 %) |
| Electron energy | 1 keV |
| Electron emission current | 20 mA |
| Current of magnetic lens | 175 mA |
| Flow rate of the gas jet | 300, 600, 1200 sccm |

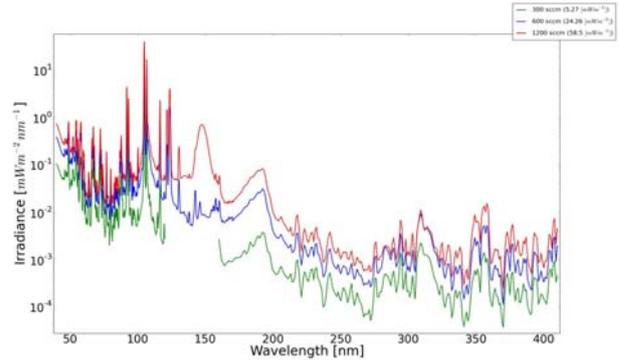

Fig. 8   Validated spectra of the VUV-simulator with an emission current of 20 mA and a gas flow of 300, 600 and 1,200 sccm.

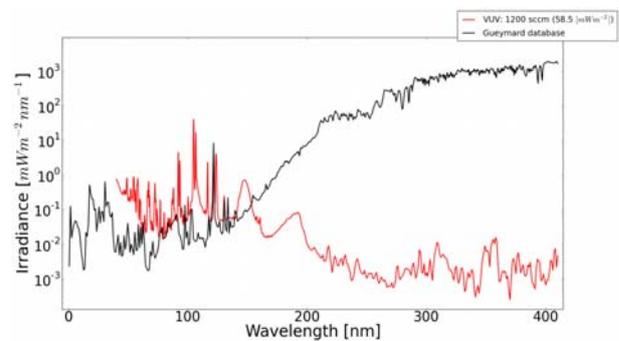

Fig. 9   Validated spectrum of the VUV-simulator with an emission current of 20 mA and a gas flow of 1200 sccm in comparison to Gueymard database [5].

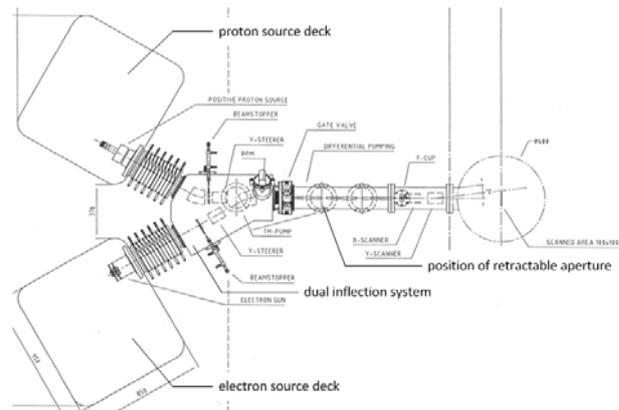

Fig. 10   Layout of the dual beam irradiation system.

The vacuum system of the beam line consists of the acceleration tubes (one for each particle species), the dual inflection system, and the differential pumping segment. The dual inflection system is equipped with a turbomolecular pump and the differential pumping segment with two ion getter pumps.

The protons are produced by ionization of Hydrogen, which is stored in a lecture bottle inside the source deck.

After pressure reducing the Hydrogen is guided through a thermo-mechanical gas inlet valve with remote control to the ion source. The ionization takes place inside the glass bulb of the source by excitation with a radio frequency, which is capacitive coupled to the bulb. The plasma is confined and positioned by an axial permanent magnetic field. The source output is optimized by control of the source gas pressure and oscillator loading. There are not only positive ions $H^+$ (1 atomic mass unit) generated with Hydrogen gas, but also the molecule ions $H_2^+$ (2 atomic mass units) and $H_3^+$ (3 atomic mass units). This makes a mass selector necessary, which is installed in the dual inflection system (Fig. 10, bottom).

The electrons are generated by a lanthanum hexaboride ($LaB_6$) cathode, which is a high performance, resistively heated, thermionic electron source. A heater current and Wehnelt voltage control the electron current.

Both particle species are accelerated in appropriate tubes by a high voltage, which corresponds to the required energy. The acceleration tubes are manufactured as a metal to ceramic brazed assembly with no organic compounds. After the acceleration the beam(s) are deflected onto a common axis in the dual inflection system. This is realized by inflection magnets. The proton inflection magnet works additionally as a mass selector for the different ion species (elemental and molecular Hydrogen ions). The electron beam is magnetically shielded from the comparatively strong magnetic fields that are applied in the proton inflection line. Both the vertical and the horizontal position of each beam are adjusted by separate magnetic steerer, which are used to compensate the influence of the electron inflection magnet to the proton beam too. The negative influence of the magnetic components of the proton beam and/or the Earth magnetic field to the electron beam is corrected with special magnetic shielding techniques, compensating magnetic fields by correction coils and advanced software tools. A separate beam stopper for each particle species can be inserted pneumatically to block the beam while the other source is running. This is useful to tune each source separately.

The beam profile monitor (BPM), Fig. 10, is used to analyze the horizontal and vertical beam size and position as well as its intensity distribution separately for each particle species. For that purpose, either the proton or the electron beam has to be blocked. The BPM is located behind the inflection segment at the beam line.

A retractable aperture can be inserted pneumatically into the beam at the differential pumping segment. It is used to reduce the current.

If the current is adjusted it can be measured with the retractable Faraday cup (F-cup, Fig. 10). It can be inserted pneumatically in front of the scanning section. This must be done separately for each particle species, i.e., again one beam must be blocked.

The electrostatic scanning segment contains two sets of deflection plates, which deflect the beam in two directions perpendicular to each other. The triangular voltages for these plates and its crystal locked frequencies are carefully chosen to eliminate the possibility of synchronization caused dose non-uniformity. The amplitude of the deflection voltages is adjusted by use of the Faraday cups, which are mounted at the corners of the sample station. They detect if the sample area is scanned completely.

## 4. Conclusions

The knowledge of degradation processes of materials under space conditions must be deepened. Therefore, the DLR Institute of Space Systems in Bremen has commissioned a new facility to study the behavior of materials under complex irradiation and to investigate their degradation in a space environment. With the CIF it is possible to irradiate samples with three light sources (for the simulation of the spectrum of solar electromagnetic radiation) simultaneously with protons and electrons. The light sources are a solar simulator with a Xenon-lamp (wavelength range 250 to 2,500 nm), a deuterium-UV-source (112 to 400 nm), and an Argon-gas-jet-VUV-simulator (40 to 410 nm). The latter enables the irradiation of samples with shorter wavelengths below the limitation of any window material. All sources have been calibrated by the PTB and DLR-Berlin. The availability of UV-radiation below 115 nm is a particular advantage of the CIF. For such small wavelengths ionization of metal atoms can appear. In the Figs. 11 and 12 the spectral distribution in the short wavelength range is presented in detail, while Fig. 13 gives an idea about the acceleration factor of the whole electromagnetic spectrum covered by the CIF sources.

As seen in Figs. 11, 12 and 13, there is a significant gap between 180 and 250 nm in the UV-spectrum and in the corresponding acceleration factor. This will be closed in the near future by modification of the solar simulator.

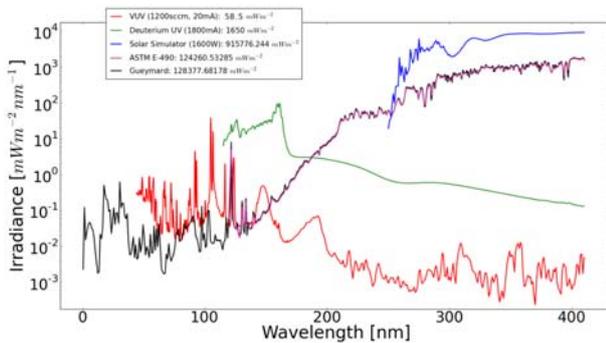

**Fig. 11  The spectra of the CIF light sources compared to the solar radiation standards [4, 5]. Both standards are shown, however, the ASTME-490 standard (pink line) reches down to 120 nm only, while the Gueymard standard (black line) coveres the range from 0.5 nm to 450 nm.**

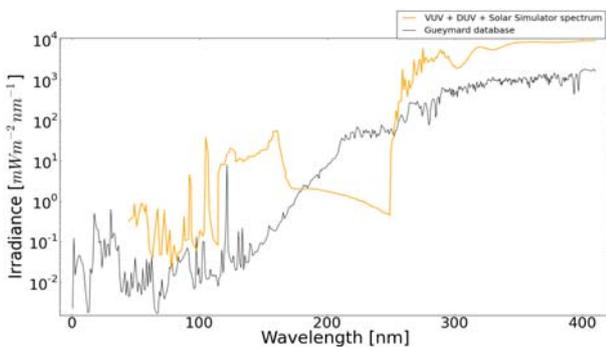

**Fig. 12  Overall spectrum of electromagnetic radiation of the CIF in the wavelength range from 40 nm to 410 nm in comparison to Gueymard database [5].**

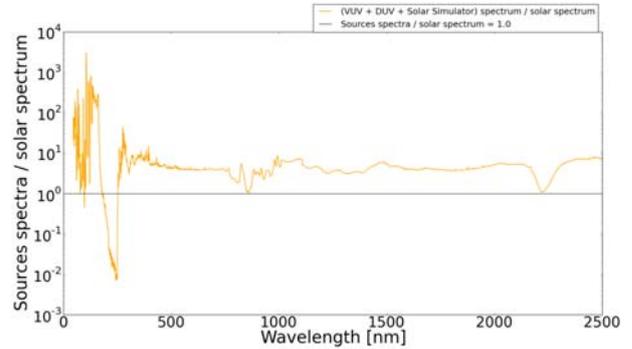

**Fig. 13  Spectral distribution of the acceleration factor of electromagnetic radiation in comparison to Refs. [4, 5] in the whole spectral range covered by the CIF radiation sources.**

In addition to the different light sources, CIF provides electron and proton irradiation. The charged particles are generated in a low energy range from 1 to 10 keV with currents from 1 to 100 nA and in a higher range from 10 to 100 keV with 0.1 to 100 μA. Both particle sources can be operated simultaneously. However, another advantage of the CIF is that electrons and protons are led in a common beam onto the sample. In order to model temperature variations as appear in free space, the sample can be cooled down to liquid Nitrogen level and heated up to about 450 °C by halogen lamps behind the target.

The complete facility has been manufactured in UHV-technology with metal sealing. It is free of organic compounds to avoid self-contamination. The different pumping systems achieve a final pressure in the $10^{-10}$ mbar range (empty irradiation chamber). Besides the installed radiation sensors, which control the stability of the various radiation sources, and an attached mass spectrometer for analyzing the outgassing processes in the irradiation chamber, the construction of CIF allows adding other in-situ measurement systems to measure parameters that are of user's interest. We are currently planning to develop an in-situ measurement system in order to determine degradation caused changes in the thermo-optical properties of the samples without leaving the vacuum.

The future activities of the CIF facility are not only focused on tests of materials which will be used in the Gossamer solar sail project [10]. The Complex

Irradiation Facility is a device that can be used in a wide spectrum of degradation investigations. Thus, it will irradiate materials that are supposed to be used in the construction of satellites such as multilayer insulations, ropes and other support structures. With its light sources together with proton and electron guns one can plan, manage and execute simulations of erosion effects of materials with large acceleration factors. Thereby, the CIF allows the study of degradation effects as may appear in long-term space missions. Moreover, the CIF will be used safely for studies in the field of planetary research.

## Acknowledgments

We like to thank Dr. Franz Lura (DLR-Berlin) and Dr. Bernd Biering (DLR-Bremen), who have initiated and escorted the design and the investment of the CIF with their wealth of experience. Additionally we like to thank Horst Schwarzer and Karl-Heinz Degen (both DLR-Berlin) for their kindly support by execution of the validation of the solar simulator and Dr. Reiner Thornagel (PTB-Berlin) for the collaboration regarding the validation of the VUV-simulator.

## References

[1] ASTM, E512-94 (2010), Standard Particle for Combined, Simulated Space Environment Testing of Thermal Control Materials with Electromagnetic and Particulate Radiation, 2010.

[2] ECSS-Q-ST-70-06C, Particle and UV Radiation Testing for Space Materials, 2008.

[3] B. Dachwald, M. Macdonald, C.R. McInnes, G. Mengali, A.A. Quarta, Impact of optical degradation on solar sail mission performance, Journal of Spacecrafts and Rockets 44 (4) (2007) 740-749.

[4] ASTM, E490-00a, Solar Constant and Zero Air Mass Solar Spectral Irradiance Tables, reapproved 2006.

[5] C.A. Gueymard, The sun's total and spectral irradiance for solar energy applications and solar radiation models, Solar Energy 76 (2004) 423-453.

[6] N. Melnik, U. Geppert, B. Biering, F. Lura, Light pressure measurment at DLR-Bremen, in: The Third International Symposium on Solar Sailing, Glasgow, 2013.

[7] M. Sznajder, U. Geppert, $H_2$-blister formation on metallic surfaces: A candidate for degradation processes in space, in: The Third International Symposium on Solar Sailing, Glasgow, 2013.

[8] M. Sznajder, T. Renger, A. Witzke, R. Thornagel, U. Geppert, Design and performance of a vacuum-UV simulator for material testing under space conditions, Advances in Space Research 52 (11) (2013) 1993-2005.

[9] E.T. Verkhovtseva, V.I. Yaremenko, Telepnev, F. Lura, Gas-jet simulator of solar VUV and soft X-ray radiation and irradiation effect on some material, in: Proceedings of the 7th International Symposium on Materials in Space Environment, Toulouse, France, 1997.

[10] U. Geppert, B. Biering, F. Lura, J. Block, M. Straubel, R. Reinhard, The 3-step DLR-ESA gossamer road to solar sailing, Advances in Space Research 48 (11) (2011) 1695-1701.